\documentclass[a4paper]{article}

\usepackage[margin=0.75in]{geometry}

\usepackage{amsfonts,amsmath,amssymb,amstext,latexsym}
\usepackage{array}
\usepackage{balance}
\usepackage{enumerate}
\usepackage{ifpdf}
\usepackage{ifthen}
\usepackage{color}
\usepackage[usenames,dvipsnames,table]{xcolor}
\usepackage{textgreek}
\usepackage{tikz}
\usepackage{xspace}

\newcommand{\thetitle}{Defining Cross-Cloud Systems}
\newcommand{\theauthor}{Yehia Elkhatib}
\date{}
\ifpdf
	\usepackage[
		colorlinks=true,%
		linkcolor=black,%
		citecolor=black,%
		filecolor=black,%
		urlcolor=black,%
		pdftex,%
		pdftitle={\thetitle},%
		pdfauthor={\theauthor},%
		pdfkeywords={cross-cloud computing, cloud computing, distributed systems, interoperability, portability, \theauthor},%
		unicode,bookmarks=false]{hyperref}
\fi

\newcommand*{\eg}{\textit{e.g.}\@\xspace}
\newcommand*{\ie}{\textit{i.e.}\@\xspace}
\makeatletter
\newcommand*{\etc}{%
    \@ifnextchar{.}%
        {etc}%
        {etc.\@\xspace}%
}
\newcommand*{\etal}{%
    \@ifnextchar{.}%
        {et al}%
        {et al.\@\xspace}%
}
\makeatother

\newboolean{showcomments}
\setboolean{showcomments}{true} 
\ifthenelse{\boolean{showcomments}}
{ \newcommand{\mynote}[3]{
    {\small$\blacktriangleright$\textsf{\textbf{\color{#3}{#2}}}$\blacktriangleleft$}}}
{ \newcommand{\mynote}[3]{}}
\newcommand{\shrink}[1]{}

\begin{document}

\title{\thetitle}

\author{
	Yehia Elkhatib\\
       School of Computing and Communications\\
       Lancaster University, UK\\
       y.elkhatib@lancaster.ac.uk
}

\maketitle

\begin{abstract}
	Recent years have seen an increasing number of cross-cloud architectures, \ie systems that span across cloud provisioning boundaries. However, the cloud computing world still lacks any standards in terms of programming interfaces, which has a knock-on effect on the costs associated with interoperability and severely limits the flexibility and portability of applications and virtual infrastructures. This paper outlines the different types of cross-cloud systems, and the associated design decisions.
\end{abstract}

\section{Introduction}
\label{sec:intro}
In cloud systems an application is developed against a remote API that is independently managed by a third party, the cloud service provider (CSP). 
A common difficulty that arises under such conditions when porting an application from consuming one set of API endpoints to another. This is born out of a number of technical and non-technical reasons. 
This added overhead of needing to adapt an application to consume a different API and switch between different contexts is sometimes minimal, but often requires a fair degree of re-engineering the application to work around the nuts and bolts of the new API. For many applications (for instance, a LAMP stack serving content to customers of a small-sized e-business), the benefit-cost ratio of interoperability is not enough to justify the required overhead. For other applications, however, it makes sense to develop them whilst incorporating interoperability aspects as a high priority design objective.

Considering the increasing realisation of the inevitability of cross-cloud computing~\cite{sotc2015,etsi-csc-wp1,Coady2015DCC}, various solutions have been proposed over recent years.
However, as expected with such an emerging field, there is a certain degree of confusion arising from the use of non-convergent terminology, such as hybrid clouds, multi-clouds, meta-cloud, federated clouds, \etc. 
The first contribution of this paper, thus, is to offer a coherent understanding of the different terminology witnessed to date in this field. 
The second contribution is a classification to capture prominent efforts in this field, describing their modus operandi and commenting on their suitability and limitations. 
These contributions are of great use to application developers and researchers in this emerging field. It also provides a model to embody the varying efforts that have and are being done, and how they relate in terms of responsibility to the different stakeholders in cloud computing.

Note that we focus on issues pertaining to interoperability at the IaaS level from the developer's perspective. These are issues primarily facing the developer of an application that transcends boundaries between different CSPs. 




\section{Why cross cloud boundaries?}
\label{sec:motivation}

We define a cross-cloud application simply as one that consumes more than one cloud API under a single version of the application. The APIs used could be dynamically interchangeable or not; this just predicates which \emph{type} of cross-cloud application it is (as introduced in the next section). However, the APIs in question have to be divergent. In other words, instrumenting an application across different data centres of one cloud service provider (CSP) is not deemed cross-cloud computing as long as such data centres expose the same API library.

A common thread to cross-cloud scenarios is change; change to the predetermined plan relating to service provisioning, use, or management.
Different parts of the application (\eg virtualized infrastructure manager, load balancer, \etc) would need to be changed to call different APIs.
Change is, of course, part of business as well as every day life. Hence, the need for cross-cloud systems grows greater as we as industries and societies use the cloud more and more.
Such change, however, is not simply binding to different service endpoints. It also entails fundamental changes to the communication behaviour in order to accommodate different semantics (\ie the interoperability question), charging model, and SLA terms. This is the core challenge of cross-cloud computing.

Another commonality is the need to be free from long-term commitment. Many consumers choose the cloud for agility and elasticity. In the past few years, this was restricted to the boundaries of a single CSP but currently the trend is to transcend different 
CSPs.
Cloud consumers do not want to be paying for resources that do not satisfy their needs and are increasingly looking for the possibility to `shop around' if they want. A recent survey~\cite{etsi-csc-wp1} carried out by the European Telecommunications Standards Institute (ETSI) discovered that the ``ability to move data from one service to another'' ranked very highly as a concern raised by private sector SMEs and large organisations that use the cloud.

As such, a number of works in both academia and industry have attempted to tackle this challenge using different strategies. Before attempting to categorize these works, it is perhaps important to point out the obvious: 
This is \textit{not} a thesis for a universally uniform provisioning system. First, such ``uber cloud'' is unrealistic given the commercial nature of the cloud computing market. Second, we believe it to be healthy to have a diverse cloud market; one where each provider brings a unique mix of specialized services that cater to a certain niche of the market. This is one aspect where the analogy of utility computing does not quite apply.

\section{Cross-cloud Dictionary}
\label{sec:types}

We identify four main types of cross-cloud systems based on the overall prevalence in the literature. We summarize these in Table~\ref{tab:types}. Our aim here is to provide some clarity about the terminology, and to use this dictionary moving forwards.
We give examples of prominent works of each type where appropriate.

\def\colwidth{3cm}
\newcolumntype{P}[1]{>{\raggedright\arraybackslash}p{#1}}
\def\arraystretch{1.2}
\newcommand{\head}[1]{\small\color{white}\textbf{#1}}

\begin{table*}[b!]
    \label{tab:types}
    \caption{The different types of cross-cloud systems.}
	\centering
	\rowcolors{2}{blue!15}{white}
    \begin{tabular}{P{2.75cm}P{\colwidth}P{\colwidth}P{\colwidth}P{\colwidth}}
    \rowcolor{black}   
    \head{}                                          & \head{Hybrid Cloud}     & \head{Multi-Cloud}                                                                                    & \head{Meta-Cloud}                                            & \head{Federated Cloud}                   \\
    \hline
    \textbf{Similarity of sub-clouds} & Dissimilar       & Similar                                                                                        & Similar                                               & Very similar, with agreement      \\
    \textbf{Level of abstraction}                          & None             & Some                                                                                           & High                                                  & Very high                         \\
    \textbf{Provisioning Means}                            & Bespoke logic    & Least common denominator API through a common programming model (CPM) or a translation library & Delegating infrastructure instrumentation             & Common API                        \\
    \textbf{Responsible Party}                             & System developer & CPM or library developer                                                                       & Third party brokers                                   & Component cloud service providers \\
    \textbf{Examples}                                      & ~                & Apache Libcloud, Apache jClouds, Fog, MODAClouds IDE                                                & OPTIMIS Toolkit, Jamcracker, Dell Cloud Marketplace & CIMI, OCCI, TOSCA, CDMI                              \\
    \end{tabular}
\end{table*}

\subsection{Hybrid Clouds}

\emph{Hybrid clouds} are those which are inherently made up of dissimilar parts~\cite{mell2011nist}. 
In a hybrid cloud system, the onus is on the system developer to amalgamate the different clouds for the purposes of their application~\cite{elkhatib2013experiences}. This also involves some logic to determine which sub-cloud is to be employed and when. Such policy logic is coupled to the rest of the application to some degree, and could be implemented as a separate module within the application or as a self-standing proxy between the application and the underlying clouds. 
In all cases, however, the policy enforcing element of the application will eventually interface with a finite set of cloud APIs. Such hardwiring is costly if the said set of APIs is to be altered after application development. However, another hidden cost seldom discussed emerges from the forward design incurred to predetermine the exact sub-clouds to be employed. This is a crucial step in order to minimize the chance of changing the set of sub-clouds after application deployment.

Although hybrid cloud architectures are abundant, they are sometimes difficult to find due to their highly customized nature which does not necessarily lend them to reuse by others in academia and especially in industry. 

\subsection{Multi-Clouds}

Like hybrid clouds, \emph{multi-cloud systems} build applications by employing different autonomous systems. However, in this model management is achieved some level of abstraction which relieves some of the cross-cloud responsibility from the developer of the end system. It also introduces a level of portability, which is a drawback of hybrid systems. There are different forms of achieving multi-clouds such as common programming models and API translation libraries. However, agreement on a least common denominator API between the different cloud systems supported by such abstraction environments does mean loss of some specialized services.

A notable example of a multi-cloud solution from academia is the mOSAIC programming model~\cite{Petcu20131417}. However, most of the traction multi-cloud engineering has had came through the use of open source API translation libraries. Despite being developed by different groups, however, these toolkits are similar in many ways. 
Libcloud is a Python toolkit that provides abstraction to more than 35 compute and data CSPs. jClouds offers an alternative for Java and Clojure developers, while Fog is a Ruby toolkit. 
Other efforts include 
DeltaCloud, and pkgcloud.

However, any need outside the unified abstract API\footnote{``One Interface To Rule Them All'', as described on the Libcloud webpage.}, \eg billing, is not supported and needs to be addressed individually by the developer. This obviously detracts from the value of such solutions.

\subsection{Meta-Clouds}

\emph{Meta-clouds}, or \emph{inter-clouds}, is another form where the responsibility is shifted away from the application developer. It differs from multi-cloud in that offers both abstraction as well as delegation. Meta-clouds are typically deployed through third party brokers that provide loosely-coupled interaction as a managed service. Such brokers handle the discovery of suitable resources and the management of the life-cycle of these resources.

Some of these brokers evolved from meta-schedulers in grid computing 
where fine-grained instrumentation of the infrastructure is out-sourced to a centralized scheduler.
An example is the work by Tordsson \etal~\cite{Tordsson2012358} on a VM-based broker that optimizes placement and management of VMs across infrastructures. 
Other efforts approach brokerage by being more engrained into the application development process. 
The Optimis Toolkit prototype presented in~\cite{Ferrer201266} constructs services as a configuration of \emph{core elements} using a programming model and runtime. Each element is attached to a set of functional and non-functional requirements that are to be honoured by a central deployment manager that negotiates the most suitable resources.

Industry offerings include Dell Cloud Marketplace, Enstratius, Jamcracker, and TDCloud.

\subsection{Cloud Federations}

A common condition in all of the above models is the lack of agreement amongst the sub-clouds. \emph{Federated clouds}, on the other hand, achieve distribution through some sort of agreement albeit minimalistic. Such cooperation could be manifested as compliance to standard interfaces and/or data formats.

To highlight one example from academia, Celesti \etal~\cite{5557976} envisioned that CSPs will federate horizontally in order to gain more market share and to further improve their capabilities by harnessing economies of scale. The Cross-Cloud Federation Manager (CCFM) is the solution they put forward to bridge the gaps between different providers, and is based on three stages: discovery, match-making and authentication. The proposed solution is to be adopted by the CSP as a middleware in order to enable them to federate with other CSPs.

Most important to note here, however, are standardisation efforts. Some of these have aimed at creating a unified interface for working with cloud resources and services across vendors. Such efforts include: 
the Cloud Infrastructure Management Interface (CIMI)~\cite{cimi-dsp0263}, 
the Open Cloud Computing Interface (OCCI)~\cite{occi-core}, 
the Topology and Orchestration Specification for Cloud Applications (TOSCA)~\cite{10.1109/MIC.2012.43}, 
and
the Cloud Data Management Interface (CDMI)~\cite{cdmi}.
In the meanwhile, other efforts have attempted to develop minimalistic APIs to support the most common interactions; \eg~\cite{Dodda09arch,6849165}.

Despite the relatively long history of standardisation groups in the cloud, such efforts are yet to pay off in terms of wide adoption by the market~\cite{ortiz2011problem}. The major players such as Amazon Web Services, Microsoft, \etc are not adopting them because they probably see such standards as giving their customers the opportunity to effortlessly switch to competitors. Instead, each CSP keeps pushing forward with their own API which is designed using their own philosophy and keep inciting customers by offering add-on services.

\section{Concluding Remarks}
\label{sec:conc}

This paper provided an overview of cross-cloud computing.
It put forward a dictionary to formalize the differences between the different solutions proposed in the literature, and gave notable examples of each. 
A lesson learned from briefly surveying such efforts is that there is no silver bullet; each solution comes with a proviso.
Further work is required in this emerging field, particularly on aspects relating to providing decision support, navigating SLA variations, and abstracting developing and shipping applications in dynamically evolving ecosystems.


%

\bibliographystyle{abbrv}
\balance{ \bibliography{crosscloud} }

\end{document}